\documentclass[aps,pra,showpacs,reprint,preprintnumbers,amsmath,amssymb,onecolumn]{revtex4-2}

\usepackage{upgreek}
\usepackage{graphicx}
\usepackage{dcolumn}
\usepackage{bm}
\usepackage{CJK}
\usepackage{multirow}

\begin{document}
	\bibliographystyle{unsrt}
	
\title{Analyzing the photoassociative spectroscopy of ultracold $\rm ^{85}Rb^{133}Cs$ in  $\rm (3)^3\Sigma^+$ state }
	\author{Zi-wei Wang}
	\affiliation{School of Physics, Dalian University of Technology, Dalian 116024, China}
	\author{Zi-ang Li}
	\affiliation{School of Physics, Dalian University of Technology, Dalian 116024, China}
	\author{Xu-hui Bai}
	\affiliation{School of Physics, Dalian University of Technology, Dalian 116024, China}
	\author{Ting Gong}
	\affiliation{State Key Laboratory of Quantum Optics and Quantum Optics Devices, Institute of Laser Spectroscopy,
		Shanxi University, Taiyuan 030006, China}
	\author{Zhong-hua Ji}
	\affiliation{State Key Laboratory of Quantum Optics and Quantum Optics Devices, Institute of Laser Spectroscopy, Shanxi University, Taiyuan 030006, China}
	\author{Yan-ting Zhao}
	\affiliation{State Key Laboratory of Quantum Optics and Quantum Optics Devices, Institute of Laser Spectroscopy, Shanxi University, Taiyuan 030006, China}
	\author{Gao-ren Wang}%
	\email{gaoren.wang@dlut.edu.cn}
	\affiliation{School of Physics, Dalian University of Technology, Dalian 116024, China}
	\date{\today}

\begin{abstract}
We establish a theoretical model to analyze the photoassociative (PA) spectroscopy of $\rm (3)^3\Sigma^+$ state of $\rm ^{85}Rb^{133}Cs$ molecule. The term energy, spin-spin coupling constant and hyperfine interaction constant of nine vibrational levels in $\rm (3)^3\Sigma^+$ state are determined. Based on the fitted term energy and rotation constant, the potential energy curve of $\rm (3)^3\Sigma^+$ state is obtained by using RKR method which is compared with the ab initial potential curve.

\end{abstract}

\maketitle
\section{\label{sec:Intro}INTRODUCTION}
Photoassociation (PA) refers to the transition from the scattering state of two atoms to a molecular excited state by absorbing a photon. PA spectroscopy in ultracold atomic gas possesses very high resolution and reveals properties of both the initial scattering state of the atom pair and the finial molecular states, which are essential for producing ultracold molecules. PA spectroscopy is also used as an sophisticated detection technique to probe the electron¨Cproton mass ratio variation \cite{gacesa2014photoassociation}, the number fluctuations in optical lattices \cite{yang2020cooling} and pair correlations in many-body systems \cite{konishi2021universal}.

PA spectroscopy in ultracold mixture of $^{85}$Rb and $^{133}$Cs atoms has been performed. Since the initial state of PA is the scattering state of the atom pair, and most of the scattering wavefunction lies in the long-range region, the Franck-Condon (FC) factor between the initial scattering state and the weakly bound excited molecular state is expected to be relatively large. Hence, the molecular levels close to the dissociation threshold of the excited electronic states are firstly detected by PA spectroscopy {kerman2004production}. Some of the weakly bound levels are identified to belong to (2)0$^{+}$ and (2)0$^{-}$ states [in Hund's case (c) notation] which are studied in details \cite{ji2011photoassociative, yuan2015investigation,su2019extensive}. Later, it is shown that the deeply bound levels in the electronically excited states can also be detected by PA spectroscopy \cite{gabbanini2011formation}, although the FC factor is expected to small. These deeply bound levels in the electronically excited states can decay spontaneously to the ground vibrational levels in ground electronic state \cite{bruzewicz2014continuous, zhao2015production, ji2017rotational}. This open up a promising pathway to continuously produce ultracold ground state RbCs molecules via PA followed by spontaneous decay. The deeply bound levels detected in \cite{gabbanini2011formation} are identified to belong to the $\Omega=0$ component of the (2)$^3\Pi$ state, where $\Omega$ is the projection of the total electronic angular momentum on the internuclear axis. Subsequently, the PA spectroscopy of this state is studied extensively \cite{ji2012photoassociative, fioretti2013experimental,bruzewicz2014continuous,yuan2015determination, zhao2016experimental}.

Recently, PA spectroscopy of the (3)$^3\Sigma^{+}$ state are performed, and several deeply bound vibrational levels, including the ground vibronic level of (3)$^3\Sigma^{+}$ state are measured \cite{shimasaki2016production}. The PA spectra of the (3)$^3\Sigma^{+}$ state have rich hyperfine structures. In contrast, no hyperfine structures are detected in the PA spectrum of the $\Omega$=0 states \cite{kerman2004production,gabbanini2011formation} due to that the leading-order hyperfine splitting vanishes. Moreover, the (3)$^3\Sigma^{+}$ state is found to be an efficient pathway to produce absolute-ground-state RbCs molecules \cite{shimasaki2016production}.

In \cite{shimasaki2016production}, the vibrational quantum numbers are assigned for the observed PA spectra. However, the hyperfine structures of the PA spectra are not studied. In this work, we establish a theoretical model to analyze the PA spectra of the $\rm ^{85}Rb^{133}Cs $ $ (3)^3\Sigma^+$ state. The hyperfine structures as well as the transition strength of the PA spectra are modeled. The coupling strengthes for the fine and hyperfine interactions $ (3)^3\Sigma^+$ are determined. Besides, we derive a Rydberg-Klein-Rees (RKR) potential energy curve of the $\rm (3)^3\Sigma^+$ state based on the spectroscopic data obtained by our model.

This article is organized as follows. In Sec.~\ref{sec:Theo}, we introduced the theoretical model used in the analysis. In Sec.~\ref{third:RES}, the model is used to study the PA spectra of $\rm ^{85}Rb^{133}Cs $ $ (3)^3\Sigma^+$ state. We summarize our work in Sec.~\ref{forth:CON}.

\maketitle
\section{\label{sec:Theo}THEORETICAL MODEL}
The experimental procedures to obtain the PA spectra are briefly summarized as follows. Initially, the pre-cooled $^{133}$Cs and $^{85}$Rb atoms are loaded in the magneto-optical trap. The PA laser is applied to associate the $^{133}$Cs-$^{85}$Rb atom pairs into molecules in excited electronic state. The excited molecules decay spontaneously to the ground electronic state. Another ionization laser is used to ionize the molecules in the ground electronic state, and the resulting molecular ions are detected. Experimentally, PA spectra is obtained by monitoring the variation of the ion signal versus the frequency of the PA laser. To analyze the PA spectra, one needs to describe properly the initial and final states coupled by the PA laser.
\subsection{INITIAL STATE}

We refer to $^{133}$Cs as atom 1, and $^{85}$Rb as atom 2. The electron spin and nuclear spin quantum numbers of ground state $^{133}$Cs atom and $^{85}$Rb atom are $s_1=1/2$, $I_1=7/2$ and $s_2=1/2$, $I_2=5/2$, respectively. The electron spin $\boldsymbol{s}_n$ and nuclear spin $\boldsymbol{I}_n$ of the same atom are coupled by hyperfine interaction to form the total angular momentum $\boldsymbol{f}_n$ where $n=1,2$. Experimentally, $^{133}$Cs and $^{85}$Rb atoms are initially prepared in the lowest hyperfine states, $f_1=3$ and $f_2=2$, respectively.
The temperature of the atomic gas is around 100~$\mu$K, and $s$-wave scattering is dominated at such ultra-low temperature. Hence, the partial wave quantum number $l$ equals to zero. The initial scattering state of the atom pair can be described by the basis $|(s_1I_1)f_1(s_2I_2)f_2flF>$, where $\boldsymbol{f}=\boldsymbol{f}_1+\boldsymbol{f}_2$, and $\boldsymbol{F}=\boldsymbol{f}+\boldsymbol{l}$. In the following, we refer to this basis as fragmentation basis. In the case considered in this work ($f_1=3$, $f_2=2$, and $l=0$), $f=F$=1,2,3,4, and 5. Initially, the states with different $F$ are assumed to coexist incoherently. Since ultracold $^{133}$Cs and $^{85}$Rb atoms are unpolarized in the experiment, the weight of the state with $F$ is $2F+1$.

\subsection{excited molecular state}
We adopt the following Hamiltonian to describe the molecules in (3)$^3\Sigma^{+}$ state
\begin{equation}
	H = H_e+H_v+H_r+H_{SS}+H_{hf},
\label{Eq:Hamiltonian}
\end{equation}
where $H_e$, $H_v$ and $H_r$ are the electronic, vibrational, and rotational hamiltonian, respectively. $H_{SS}$ describes the spin-spin interaction, and $H_{hf}$ is the hyperfine interaction.
In the following, we adopt Hund's case b basis to obtain the matrix representation of $H$. Conventionally, the Hund's case b basis is expressed as $|\Lambda S \nu N J>$, where $\Lambda$ is the projection of the total electronic orbital angular momentum on the internuclear axis, $\boldsymbol{S}=\boldsymbol{s_1}+\boldsymbol{s_2}$ is the total electronic spin angular momentum. For (3)$^{3}\Sigma^+$ state, $\Lambda$ equals to 0, and $S$ is 1. $\nu$ is the vibrational quantum number. $\boldsymbol{N}$ results from the coupling between $\boldsymbol{\Lambda}$ and the rotational angular momentum $\boldsymbol{R}$. Since $\Lambda$ is zero for (3)$^{3}\Sigma^+$ state, $\boldsymbol{N}$ is essentially the rotational angular momentum. As only $s$-wave scattering state is considered initially, the quantum number $N$ is fixed to be 1 due to transition selection rule. $\boldsymbol{J}=\boldsymbol{N}+\boldsymbol{S}$ is the total angular momentum with the nuclear spin angular momentum excluded. In this work, we extend the conventional Hund's case b basis to include the nuclear spin angular momentum. The coupling scheme is as follows. $\boldsymbol{J}$ is firstly coupled with the nuclear spin angular momentum of $^{133}$Cs $\boldsymbol{I_1}$ to form $\boldsymbol{F_1}$, and then $\boldsymbol{F_1}$ is coupled with the nuclear spin angular momentum of $^{85}$Rb $\boldsymbol{I_2}$ to form $\boldsymbol{F}$. The full basis used in this work is expressed as $|\Lambda S \nu N J I_1 F_1 I_2 F>$. Since we focus on the (3)$^3\Sigma^+$ state in this subsection, we will use the notation $|^3\Sigma^+\nu N J I_1 F_1 I_2 F>$.

The first three terms in Eq.~(\ref{Eq:Hamiltonian}) $H_e+H_{\nu}+H_r$ is diagonal matrix in Hund's case b basis, and the diagonal matrix element is given by
\begin{equation}
<^3\Sigma^+{\nu}NJ|H_e+H_{\nu}+H_r|^3\Sigma^+{\nu}NJ>=E_{\nu}+B_{\nu}N(N+1),
\label{Eq:matrixElementHeHvHr}
\end{equation}
where $E_{\nu}$ and $B_{\nu}$ are the energy and rotational constant of the vibrational level $\nu$. The diagonal matrix elements of spin-spin interaction are as follows
\begin{equation}
	<^3\Sigma^{+} \nu N J |H_{SS}|^3\Sigma^{+} \nu N J > = \left\{
    \begin{array}{lll}
    - \frac{N+1}{2N-1} \times \frac{2}{3} \lambda_{\nu},   &   & {J=N-1,} \\
    \frac{2}{3} \lambda_{\nu},                             &   & {J=N,}      \\
    - \frac{N-1}{2N+3} \times \frac{2}{3} \lambda_{\nu},   &   & {J=N+1,}    \\
    \end{array} \right.
\label{Eq:diagonalMatrixElementHss}
\end{equation}
where $\lambda_{\nu}$ represents the spin-spin coupling constant of the vibrational level $\nu$. The nonvanishing off-diagonal matrix elements are expressed by
\begin{equation}
\begin{array}{lllll}
<^3\Sigma^{+} \nu N J |H_{SS}|^3\Sigma^{+} \nu N^{\prime} J >=
\frac{3\sqrt{(N+1)(N+2)}}{2N+3} \times \frac{2}{3} \lambda_{\nu}, & &  {N^{\prime}=N+2,} && {J=N+1.}
\end{array}
\label{Eq:offdiagonalMatrixElementHss}
\end{equation}
The hyperfine interaction includes two parts. The hyperfine interaction between the nuclear spin of $^{133}$Cs atom and the total electronic spin is denoted by $H_{hf}(1)$, and the hyperfine interaction related to the nuclear spin of $^{85}$Rb atom is denoted by $H_{hf}(2)$. The diagonal matrix elements of hyperfine interaction $H_{hf}(1)$ are given by
\begin{equation}
	<^3\Sigma^{+} \nu N J I_1 F_1 I_2 F|H_{hf}(1)|^3\Sigma^{+} \nu N J I_1 F_1 I_2 F> = \left\{
    \begin{array}{llll}
    - K_{1}G(F_1,N-1,I_1)/N   &   & {J=N-1,} \\
    K_{1}G(F_1,N,I_1)/N(N+1), &   & {J=N,}      \\
    K_{1}G(F_1,N+1,I_1)/N+1,      &   & {J=N+1,}    \\
    \end{array} \right.
\label{Eq:diagonalMatrixElementHhf1}
\end{equation}
where $K_1$ is the hyperfine interaction constant which characterizes the strength of $H_{hf}(1)$, and $G(F,N,I)=[F(F+1)-N(N+1)-I(I+1)]/2$. The nonvanishing off-diagonal matrix elements of $H_{hf}(1)$ is
\begin{equation}
\begin{aligned}
	&<^3\Sigma^{+} \nu N J I_1 F_1 I_2 F|H_{hf}(1)|^3\Sigma^{+} \nu N J^{\prime} I_1 F_1 I_2 F>  \\
    &= \left\{
    \begin{array}{lllll}
    - K_{1} \frac{1}{2N}\left(\frac{N+1}{2N+1}\right)^{1/2}
    [(F_1+N+I_1+1)(N+I_1-F_1) &&&& \\
     \quad\times(F_1+N-I_1)(F_1+I_1+1-N)]^{1/2},   &   & {J=N,} && {J^{\prime}=N-1}, \\
    -K_{1}\frac{1}{2(N+1)}\left(\frac{N}{2N+1}\right)^{1/2}[(F_1+N+I_1+2)(N+I_1+1-F_1) &&&& \\
    \quad\times(F_1+N+1-I_1)(F_1+I_1-N)]^{1/2}, &   & {J=N,} && {J^{\prime}=N+1}. \\
    \end{array} \right.
\end{aligned}
\label{Eq:diagonalMatrixElementHhf1}
\end{equation}
The diagonal matrix elements of $H_{hf}(2)$ are expressed as
\begin{equation}
\begin{aligned}
    &<^3\Sigma^{+} \nu N J I_1 F_1 I_2 F|H_{hf}(2)|^3\Sigma^{+} \nu N J I_1 F_1 I_2 F>  \\
    &= \left\{
    \begin{array}{llll}
    K_2{G(I_1,F_1,N-1)G(F,F_1,I_2)}/[{NF_1(F_1+1)}],   &   & {J=N-1,} \\
    -K_2{G(I_1,F_1,N)G(F,F_1,I_2)}/[{N(N+1)F_1(F_1+1)}], &   & {J=N,}      \\
    -K_2{G(I_1,F_1,N+1)G(F,F_1,I_2)}/[{(N+1)F_1(F_1+1)}],      &   & {J=N+1,}    \\
    \end{array} \right.
\end{aligned}
\label{Eq:diagonalMatrixElementHhf2}
\end{equation}
where $K_2$ is the hyperfine interaction constant characterizing the interaction strength between the nuclear spin of $^{85}$Rb and the electronic spin. Since $H_{hf}(2)$ is weak, the off-diagonal matrix elements of $H_{hf}(2)$ are neglected.

The total angular momentum $F$ is a good quantum number. The molecular states with $F=0,1,2,3,4,5,6$ can be excited from the initial scattering state due to transition selection rule. For each $F$, we diagonalize the corresponding Hamiltonian matrix to obtain the energy levels and eigenstates.
\subsection{Line strength}
To characterize the transition probability between the initial scattering state and the final excited molecular state, we calculate the line strength which is the square of the transition moment. In the following, we firstly expand the fragmentation basis in the Hund's case (b) basis, and then present the expression of the transition moment in Hund's case (b) basis.

Since the two atoms are in ground state initially, the total electronic orbital angular momentum $\boldsymbol{L}$ and its projection on the internuclear axis $\Lambda$ equal to 0. As a result, $N$ in the Hund's case (b) basis is identical with $l$ in the fragmentation basis, which describes the rotation between the two nuclei. Since $s$-wave scattering state is considered, $N$ is 0. Furthermore, $J$ is the same with $S$ due to $N=0$. Under such circumstance, the fragmentation basis can be expressed as
\begin{equation}
	\begin{split}
		|(s_1I_1)f_1(s_2I_2)f_2flF>=&\sum_{SIJF_1}(-1)^{l+S+J+I_1+I_2+I+2F}
        [(2f_1+1)(2f_2+1)(2f+1)(2S+1)(2J+1)(2F_1+1)]^{1/2}(2I+1) \\
        &\times\begin{Bmatrix}
			S&I_1&F_1\\
			I_2&F&I
		\end{Bmatrix}
        \begin{Bmatrix}
			s_1&I_1&f_1\\
            s_2&I_2&f_2 \\
			S&I&f
		\end{Bmatrix}
|\Lambda S N J I_1 F_1 I_2 F>  ,
	\end{split}
\label{Eq:matrixElementTransitionDipoleMoment}
\end{equation}

where $\boldsymbol{I}=\boldsymbol{i_1}+\boldsymbol{i_2}$ is the total nuclear spin angular momentum, and the two-row and three-row arrays in the curly bracket are Wigner 6$j$ and 9$j$ symbols, respectively.

The expression of transition moment in Hund's case (b) basis excluding the nuclear spin is derived In Ref \cite{watson2008hon}. In the Hund's case (b) basis adopted in this work, the expression of the transition moment is modified to be
\begin{equation}
	\begin{split}
		&<\Lambda S N J I_1 F_1 I_2 F| \boldsymbol{d} |\Lambda' S' N' J' I_1 F_1' I_2 F'>  \\
     &= \delta_{SS'}(-1)^{p}[(2F+1)(2F'+1)(2F_1+1)(2F'_1+1)(2J+1)(2J'+1)(2N+1)(2N'+1)]^{\frac{1}{2}} \\
        &\quad\times
		\begin{pmatrix}
			N&1&N'\\
			-\Lambda&\Lambda'-\Lambda&\Lambda'
		\end{pmatrix}
		\begin{Bmatrix}
			N&1&N'\\
			J'&S&J
		\end{Bmatrix}
        \begin{Bmatrix}
			J&I_1&F_1\\
			F_1'&1&J'
		\end{Bmatrix}
        \begin{Bmatrix}
			F_1&I_2&F\\
			F'&1&F_1'
		\end{Bmatrix}
		M_{\Lambda-\Lambda^{\prime}},
	\end{split}
\label{Eq:matrixElementTransitionDipoleMoment}
\end{equation}
where $p=F'+F_1+F_1'+J+J'+I_1+I_2+S+1-\Lambda$,  $M_{\Lambda-\Lambda^{\prime}}=<\Lambda|d_{\Lambda-\Lambda^{\prime}}|\Lambda^{\prime}>$, and $d_{\lambda}$ is the $\lambda=0,\pm{1}$ component of the electric dipole moment $\boldsymbol{d}$.  As stated above, both $\Lambda$ and $\Lambda^{\prime}$ for the levels before and after the PA transition considered in this work are 0. Therefore, $M_{0}$ component is relevant in our calculation. Furthermore, we assume $M_{0}$ is constant in this work. Since the relative transition probability is considered, the value of $M_{0}$ is set to be 1. In Eq.~(\ref{Eq:matrixElementTransitionDipoleMoment}), the array in the parentheses is winger 3$j$ symbol.

\section{\label{third:RES}RESULTS AND DISCUSSIONS}
\begin{figure*}[t!]
	\includegraphics[width=0.9\textwidth]{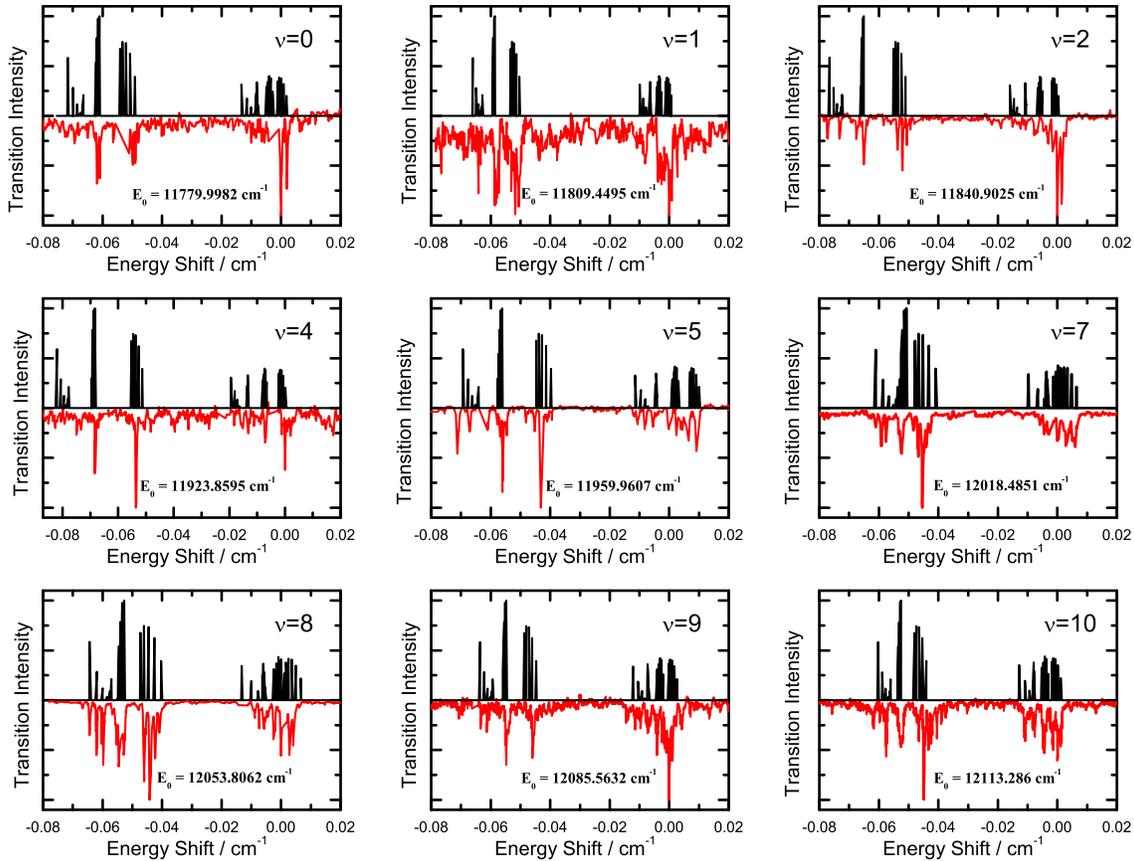}
	\caption{\label{Fig:1}
		Observed (red line) and calculated (black line) hyperfine splittings of vibrational levels for $\nu = 0 \ (a), \nu = 1 \ (b), \nu = 2 \ (c), \nu = 4 \ (d), \nu = 5 \ (e), \nu = 7 \ (f), \nu = 8 \ (g), \nu = 9 \ (h) $ and $ \nu = 10 \ (i)$. The ordinate is the relative transition intensity, which is characterized by the molecular ionization intensity in the experiment, and theoretically corresponds to the relative transition dipole moment. The abscissa is the relative shift of the energy levels ,the unit is the wave number $\rm (cm^{-1})$. The most dominant peak in the second group of peaks is selected as the zero point. }
\end{figure*}

In this section, the PA spectra of the $^{85}$Rb$^{133}$Cs (3)$^3\Sigma^+$ state are analyzed. In the analysis, the
vibrational energy $E_\nu$, the spin-spin coupling constant $\lambda$, and the hyperfine interaction constant $K_1$ and $K_2$ are treated as fitting parameters. The rotational constant $B_\nu$ is calculated with the potential obtained in \cite{lim2006theoretical}. We show the comparison between the experimental and theoretical spectra for vibrational levels $\nu = 0-2, 4, 5, 7-10$ in Fig.~\ref{Fig:1}. The values of fitting parameters for the observed levels are listed in Table ~\ref{Table:Para}.

Since the rotational quantum number of the initial scattering state is $l = 0$, the energy levels of the excited state
are mainly composed of $N = 1$. For the $(3)^3\Sigma$ electronic state, the molecular electron spin angular momentum $S = 1$,
so the quantum number $J = 0, 1, 2$. The two sets of peaks observed experimentally are identified to be $J = 1$ and $2$,
respectively. In the Hamiltonian model we established above, $F$ is a good quantum number. The spin-spin interaction
couples different rotational quantum numbers $N$ of the same $F$, and the degeneracy of $J$ is eliminated. Spin-spin
interaction is the leading origin of energy level splittings. The further split is due to the hyperfine interaction of Rb and Cs atoms. The split produced by the hyperfine interaction is two orders of magnitude smaller than the spin-spin
interaction.

\begin{figure}[t!]
	\includegraphics[width=0.45 \textwidth]{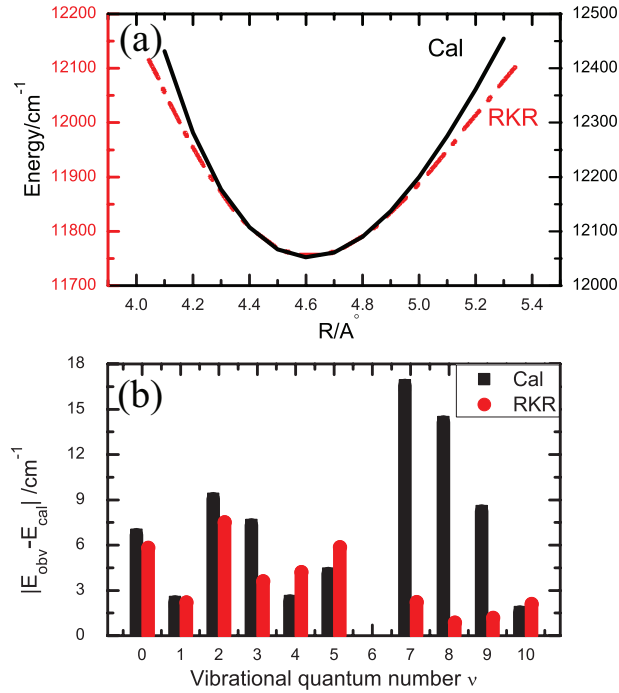}
	\caption{\label{Fig:2}
		The potential energy curve of $(3)^3\Sigma$ state calculated by Lim et al.\cite{lim2006theoretical}(black line, vertical axis on the right) and by the RKR method in our work (red dash-dot line,vertical axis on the left). (b) The absolute error (under the same potential
		energy reference point) of Lim's potential energy curve (black line) and the
		RKR potential energy curve (red line). }
\end{figure}

According to the vibrational energy $E_\nu$ of $^{85}$Rb$^{133}$Cs $(3)^3\Sigma^+$ state (as shown in Table ~\ref{Table:Para}), we use the RKR program\cite{le2017rkr1} to refine the potential energy curve (as shown in Fig.~\ref{Fig:2}(a)). The red dotted-dashed line is the potential curve reconstructed in our work through the RKR program, and the black solid line was obtained via ab initio
calculations in \cite{lim2006theoretical}. We use the second-order Dunham function to fit the observations, the vibrational constant and
anharmonic constant are $\omega _e = 38.259$ $\rm cm^{-1}$ and $\omega _e \chi _e = 0.38$ $\rm cm^{-1}$ , respectively. We reasonably translate the two
potential energy curves to the appropriate energy reference point and input them into the level program\cite{le2017level} to solve
the Schrodinger equation. Taking the absolute difference between the solved eigenvalue and the observed energy
levels, we get the accuracy of the two potential energy curves. As shown in Fig.~\ref{Fig:2}(b), the red bar represents the
absolute error between the eigenvalue of the potential energy curve constructed by RKR and the observed value, and
the black bar is that for Lim et al. It can be seen that the potential energy curve in our work accords with the experiment better. The sum of errors is only about half of the potential energy curve obtained by Lim et al., and the
error fluctuation range is smaller.

\section{\label{forth:CON}CONCLUSIONS}
We established a theoretical model under the 5s+6p dissociation limit to analyze the hyperfine structures of the $\rm ^{85}Rb^{133}Cs$ molecular $(3)^3\Sigma$ state. The analysis includes the relative splittings and the relative transition intensity of each branch, and the experimental observation results are reproduced accurately. It is found that the hyperfine splittings are mainly derived from spin-spin interaction and hyperfine interaction. We have determined the parameters of these effects. Based on experimental observations, we have used the RKR method to reconstruct the potential energy curve of $\rm ^{85}Rb^{133}Cs$ molecular $(3)^3\Sigma$ state, which has significantly improved the accuracy compared to the previous potential energy curve.

\begin{table*}[t!]
	\caption{\label{Table:Para} Term energy $E_\nu$, rotation constant $B_\nu$, spin-spin interaction parameters $\lambda$, hyperfine interaction parameters for Cs atoms $K_1$ and Rb atoms $K_2$ for the $\rm ^{85}Rb^{133}Cs \ (3)^3\Sigma^+$ state $v = 0-2, 4, 5, 7-10$ are summarized. We consider the energy difference between the theoretically obtained relative hyperfine splittings and the experimentally observed values as the term energy $E_\nu$. And rotation constant $B_\nu$ are calculated from the potential we refined. }
	\begin{ruledtabular}
		\begin{tabular}{l c c c c c c c c c }
			$\nu$ & 0 & 1 & 2 & 4 & 5 & 7 & 8 & 9 & 10
			\\ \hline
			$E_\nu \ \rm{(cm^{-1})}$ &
			11780.28
			&11809.66
			&11841.31
			&11924.40
			&11960.21
			&12018.60
			&12053.92
			&12085.68
			&12113.35
			\\
			$B_\nu \ \rm{(cm^{-1})}$ & 0.01529&0.01526&0.01523&0.01515&0.01511&0.0150&0.01496&0.0149&0.01484\\
			$\lambda \quad \rm{(cm^{-1})}$ & -0.55&-0.45&-0.75&-0.95&-0.5&-0.29&-0.29&-0.3&-0.22 \\	
			$k_1 \  \ \rm{(cm^{-1})}$  & 0.005&0.0035&0.006&0.004755&0.005&0.004&0.0051&0.0042&0.0035\\
			$k_2 \ \ \rm{(cm^{-1})}$ &0.0025&0.0015&0.002&0.002&0.0025&0.0035&0.0035&0.002&0.002\\
			
		\end{tabular}
	\end{ruledtabular}	
\end{table*}

The PA spectrum of $v=3$ vibrational level has also probed experimentally, the pattern of which is different from those of the vibrational levels considered in this work. As pointed out in \cite{shimasaki2016production}, this is probably due to that there is a vibrational levels of the (2)$^{3}\Pi_{0^-}$ state very close to the $v=3$ vibrational level, and the coupling between these levels results in particular pattern of the PA spectrum. In the future, we will expand the current model to include the (2)$^{3}\Pi_{0^-}$ state, and analyze the PA spectrum of $v=3$ level.
\section{\label{forth:ACK}ACKNOWLEDGMENTS}
This work is supported by the National Key R$\&$D Program of China No.2018YFA0306503, the National Natural Science Foundation of China under Grant No. 12104082, and the Fundamental Research Funds for the Central Universities DUT19LK35.

\bibliography{ref}
\end{document}